\documentclass{aa}
\usepackage[dvips,xdvi]{graphics}
\def\HA{\mbox{H$\alpha$}}
\def\HB{\mbox{H$\beta$}}

\def\mic{\mbox{$\mu$m}}

\def\L{\mbox{$\lambda$}}

\begin{document}

\title{
ISO--SWS spectroscopy of IC443 and the origin of the IRAS 12 and 25 $\mu$m
emission from radiative supernova remnants
\thanks{ Based on observations with ISO, an ESA project
with instruments funded by ESA Member States (especially the PI countries:
France, Germany, the Netherlands and the United Kingdom)
and with the participation of ISAS and NASA. 
The SWS is a joint project of SRON and MPE.
}}

\author{E. Oliva\inst{1}, 
D. Lutz\inst{2}, 
S. Drapatz\inst{2},
and
A.F.M. Moorwood\inst{3},
}

\institute{
Osservatorio Astrofisico di Arcetri, Largo E. Fermi 5, 
I--50125 Firenze, Italy
\and
Max Planck Institute f\"ur Extraterrestrische Physik,
            Postfach 1603, D-85740 Garching, Germany
\and
European Southern Observatory, Karl Schwarzschild Str. 2, D-85748
Garching bei M\"unchen, Federal Republic of Germany 
}

\offprints{E. Oliva}

\date{Received 17 November, accepted 3 December 1998 }

\titlerunning{SWS spectroscopy of IC443 and the origin of the
blue IRAS emission from 
radiative SNRs}
\authorrunning{Oliva et al. }
\thesaurus{01(                  % Section Letters
               09.09.1 IC443;    % ISM: individual objects
               09.19.2;          % ISM: supernova remnants
	       13.09.4           % Infrared: ISM: lines and bands
             )}
\maketitle

\begin{abstract}

ISO--SWS spectral observations
of the supernova remnant IC443 are presented. Like other
radiative SNRs, this object
is characterized by prominent line--emitting filaments
and relatively strong IRAS 12 and 25 \mic\   emission which is
commonly interpreted as thermal radiation from very small gra\-ins 
stochastically heated by collisions with the hot
plasma behind the shock front.
This interpretation is challenged by 
the data presented in this {\it Letter} which indicate that most
of the 12 and 25 \mic\  IRAS flux 
is accounted for by ionized line emission (mainly [NeII] and [FeII]).
This result also seems to hold for other radiative SNRs.
We also discuss
the  possible contribution of H$_2$ lines to the IRAS 12 \mic\
flux from the southern rim of IC443
and briefly analyze the element abundances derived from the observed
ionic lines.

\end{abstract}

\section{Introduction}

A significant fraction of the galactic supernova remnants were
detected by the IRAS satellite (e.g. Arendt 1989, hereafter \cite{A89}) and
their relatively strong FIR emission has been 
normally interpreted as thermal radiation from dust which is
heated by collisions with the hot (million degrees) plasma
in the post--shock region (e.g. Braun \& Strom \cite{braun86},
Dwek et al. \cite{dwek_etal87}). This emission could be an important
cooling mechanism for non--radiative shocks 
in young SNRs (e.g. Ostriker \& Silk \cite{ostriker73}).

The spectral shape of the FIR emission reflects the temperature distribution
of the dust which in turn depends on the shock speed and on the
density, size and composition of 
the grains (e.g. Draine \cite{draine91}). 
Young SNRs are expected to have a FIR spectrum characterized by a single
temperature because, in the hot ($\ga\!2\,10^7$ K) plasma typical 
of these objects, the temperatures of collisionally heated
dust grains approach the same value independent of grain size 
(e.g. Dwek \cite{dwek87}). This is in good agreement with the IRAS 
data of young SNRs (e.g. Cas A, Tycho, Kepler) whose spectral distributions
are fairly well fitted by a single temperature blackbody (e.g.  \cite{A89})

\begin{figure*}
\def\AST{\mbox{\Large\bf{$\ast$}}}
\centerline{\resizebox{\hsize}{!}{\rotatebox{0}{\includegraphics{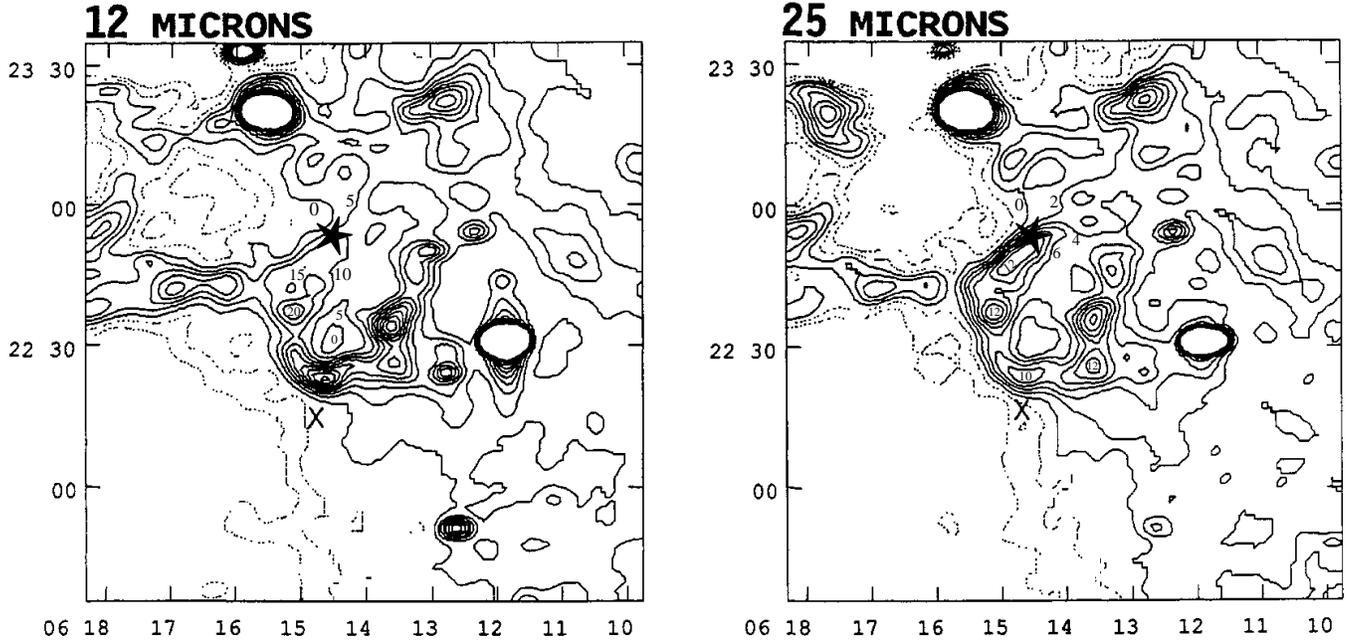}}}}
%\vskip-5pt
\caption{
The position of the SWS aperture ($\AST$) is overlaid onto a
reproduction of the IRAS maps
of IC443 (adapted from Fig. 30a \cite{A89}).
The solid contours correspond to 0, 5, 10, 15, 20, 25, 30, 35, 40, 45
and 0, 2, 4, 6, 8, 10, 12, 14, 16, 18 
($\times10^{-8}$ W m$^{-2}$ sr$^{-1}$) in the 12 and 25 \mic\  bands,
respectively.
%
% Added
%
Note that the size of the SWS aperture is much smaller than the symbol used
to mark its position.
%
% In iras_map.fig 1arcmin=87.5 units
%
}
\label{iras_map}
\end{figure*}

The IRAS colours of radiative SNRs are much more complex
and require an ad hoc combination of dust components of different sizes and
temperatures. In particular, the S(12)/S(25) IRAS colours are much bluer
than those of young SNRs and require a very large population of 
very small ($<\!50$ \AA) grains (e.g. Arendt et al. \cite{arendt92}).
An alternative explanation is that the flux seen in the
``blue'' IRAS bands is dominated by line emission from the radiative
filaments but this possibility could not be verified directly because of the
difficulty of measuring the FIR lines prior to ISO. Attempts
to estimate the line contamination by extrapolating optical spectral
measurements using
shock model predictions have led to contradictory conclusions
(e.g. Mufson et al. \cite{mufson}, Arendt et al. \cite{arendt92}).

In this {\it Letter} we present FIR spectral
observations of a prototype radiative SNR which allow a direct comparison
between the IRAS and line fluxes.
The data are presented in Sect.~\ref{data}, the results are discussed
in Sect.~\ref{discussion} and
in Sect.~\ref{conclusions} we draw our conclusions.

\section { Observations and results }
\label{data}

A complete SWS01 (speed 4, 6700 sec total integration time) spectrum
centered at $\alpha$=06$^h$14$^m$32$^s$.6  $\delta$=22$^o$54'05''
(1950 coordinates) 
%
% added
%
and roughly
corresponding to position 1 of Fesen \& Kirshner (\cite{fesen80})
was obtained on October 16, 1997.
The data were reduced using standard routines of the SWS interactive
analysis system (IA) using calibration tables as of October 1998.
Reduction relied mainly on the default pipeline steps, plus
removal of signal spikes, elimination of the most noisy
band 3 detectors, and flat--fielding.
%
% Modified
%
Note that SWS has a quite high spectral resolution
($R\!=\!\lambda/\delta\lambda\!\simeq\!1500$) and is not  well
suited for measurements of faint continuum fluxes. For the SWS01 mode, 
in particular, s/n and detector drifts 
between dark current measurements do not allow measurements of continuum 
emission at levels much below 1 Jy.
%
% 1e-7 W m-2 sr-1 => 7.4e5 Jy/sr at 12 mic => 0.007 Jy in 14x27 sqarcsec
%
The measured IRAS 12 and 25 \mic\   ``continuum'' fluxes from the SNR 
would correspond to only $\simeq$0.01 Jy in the SWS aperture
and cannot therefore be detected
with the spectral observations presented here.

\begin{figure}
\centerline{\resizebox{\hsize}{!}{\rotatebox{000}{\includegraphics{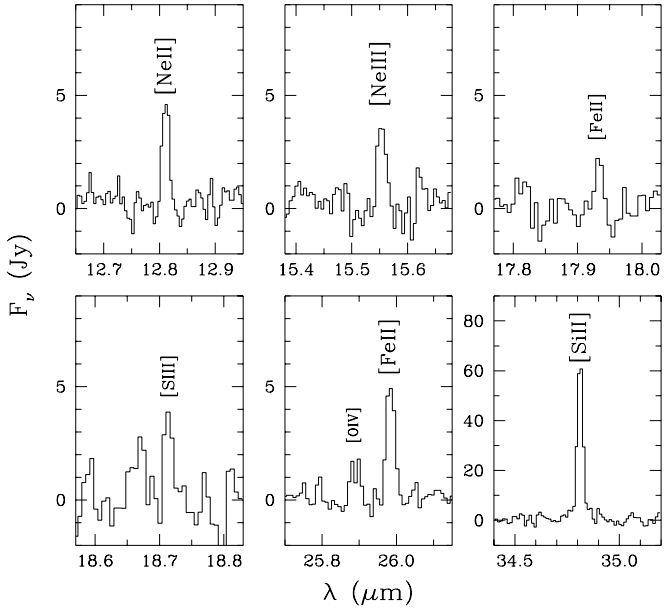}}}}
%\vskip-5pt
\caption{
Selected sections of the SWS01 spectrum of the north--east rim
of IC443, note that each spectral segment has been ``continuum subtracted''
(using a $\le$2 degree polynomial) to remove instrumental offsets and drifts.
The position of instrument aperture
is shown in Fig.~\ref{iras_map}.
}
\label{sws01}
\end{figure}

The spectral sections including well detected lines are displayed in
Fig.~\ref{sws01} and the derived line fluxes
are listed in Table~\ref{tab_flux} together with their contribution
to the IRAS fluxes, estimated using the spectral response shown
in Fig.~\ref{iras_filt}.
%
% Added
%
The line profiles are not resolved within the noise (i.e. FWHM$<$400 km/s),
and their centroids are not significantly red/blue--shifted.

The most striking result is that the surface brightness of [NeII]\L12.8
alone is {\it larger} than that observed by IRAS through  the
12 \mic\  filter.
This apparent contradiction reflects the somewhat higher spatial resolution
of ISO--SWS relative to the undersampled IRAS maps. 
%
% Modified
%
More specifically, the line emitting filaments visible in the \HA\  
and [SII] images of Mufson et al. (\cite{mufson}) are quite
uniformly distributed over many arcmin, i.e. an area larger than the IRAS 
map resolution, and the average line surface brightness over this area
is roughly a factor of 2--4 lower than that within the ISO--SWS aperture.
In other words, the ISO spectra, although centered on a bright optical
filament, does not sample a spot of exceptionally
large line surface brightness, but rather a ``typical'' line emitting region.

This indicates, therefore, that the IRAS 12 \mic\  ``continuum'' is indeed 
dominated by [NeII] line emission, i.e. that this line accounts
for at least 50\% of the observed IRAS flux.
A similar reasoning applies to the IRAS 25 \mic\  flux which is strongly
contaminated by emission in the [FeII]\L26.0 ground state transition
plus significant contribution from [OIV]\L25.9, [SIII]\L18.7 and 
[FeII]\L17.9 (cf. Table~\ref{tab_flux}).\\

%
% Fluxes relative to [FeII]26.0 in IC443 and RCW103
%       [NeII]12.8                  2.2       1.33
%       [FeII]17.9                  0.6       0.57
%       [SIII]18.7                  0.8       0.39
%       [OIV]25.9                   0.4       0.30
%       [SiII]34.8                  6.5       3.1   asuming a point source
%       [SiII]34.8                  3.7       2.1   RCW103 OK, IC443 uniform
%

Another interesting result is that [FeII]\L26.0/[NeII]\L12.8
and [FeII]\L26.0/[SiII]\L34.8 
are both a factor of $\simeq$1.7
lower than those observed in RCW103 (Oliva et al. \cite{rcw103_iso}),
while the density sensitive [FeII]\L17.9/\L26.0 ratio is similar
in the two objects. 
This result indicates that IC443 has a lower Fe gas phase abundance than RCW103 
where the Fe/Ne and Fe/Si relative abundances 
were found to be close to their solar values. 
%
% Modified
%
This difference may reflect intrinsic differences in the ISM total element
abundances or, more likely, imply that the shock in IC443 
is slower and therefore less effective in destroying the Fe--bearing grains. 
This last possibility is also supported by other ``speedometers'',
i.e. the line surface brightness and the
[NeIII]/[NeII] ratio which are lower by a factor of 7 and 1.6, 
respectively.

\section{ Discussion }
\label{discussion}

%
% Modifyed
%
The data presented above strongly suggest that most of the 12 and 25 \mic\
flux from the NE rim of IC443 is due to ionic line emission rather than
continuum emission from warm dust.
We address here the following questions.
Can this result be generalized to other radiative SNRs?
Are there other lines which may severely contaminate IRAS measurements
of this and other remnants?

\subsection{The IRAS 12 and 25 \mic\  emission from other SNRs}

%
% Added
%
The easiest and most direct estimate of the line contribution to the
IRAS fluxes requires a measurement of the FIR line intensity
from the whole SNR.
This is virtually impossible in IC443 and other radiative
remnants in the Galaxy because of their very large projected sizes, but is
feasible in the LMC where e.g. the remnant N49 has been mapped with SWS
by Oliva et al. (in preparation). They find a total [NeII]\L12.8  line 
intensity
of $2\,10^{-14}$ W m$^{-2}$ and equal, within the errors, to the
F(12\mic)=$2.2\,10^{-14}$  W m$^{-2}$
IRAS flux reported by Graham et al. (\cite{graham87}).
 
%
% in RCW103 [NeII]/Hbeta~1.1, [FeII]26.0/Hbeta~0.85
%
For remnants where direct measurements of FIR lines is not available,
a reasonable estimate of their fluxes
can be obtained by scaling optical line measurements using 
available ISO spectral observations of radiative SNRs, namely
IC443 (this paper),
RCW103 (Oliva et al. \cite{rcw103_iso}) and N49 (Oliva et al.
in preparation). These indicate that the [NeII] and [FeII]+[SIII]+[OIV]
line contribution to
the 12 and 25 \mic\  filters are both roughly equal to the flux of \HB.
Unfortunately, relatively few optical spectrophotometric observations of SNRs
are available in the literature and are altogether missing for several
important radiative remnants such as W44 and W49B.
Nevertheless, the SW filament of RCW86 has
$\Sigma(\HB)\!\simeq\!4\,10^{-7}$ 
(Leibowitz \& Danziger \cite{leibowitz}) and very similar
to the $\Sigma\!\simeq\!3\,10^{-7}$ W m$^{-2}$ sr$^{-1}$ IRAS 12 and 25
\mic\   fluxes 
found by \cite{A89}. Similarly, the NE filament of the Cygnus Loop 
has $\Sigma(\HB)\!\simeq\!1\,10^{-7}$ (Fesen et al. \cite{fesen82})
and close to the IRAS surface brightness
$\Sigma\!\simeq\!8\,10^{-8}$ W m$^{-2}$ sr$^{-1}$ (\cite{A89}).
These results suggest, therefore, that lines account for most of
the IRAS 12 and 25 \mic\  emission
from line emitting filaments of radiative SNRs.

Another interesting exercise is to compare optical line and IRAS fluxes
in a younger remnant such as the Kepler SNR for which accurate line
photometric measurements are available (D'Odorico et al. \cite{dodorico}).
The total H$\beta$ emission from the whole remnant is $2.9\,10^{-15}$
W m$^{-2}$ and only 2\% and 0.6 \% of the IRAS flux in the 12 and 25
\mic\  bands, respectively. This indicates, therefore, that line 
contamination is negligible in this object.

\subsection{ Contamination from other lines }

\begin{figure}
\centerline{\resizebox{\hsize}{!}{\rotatebox{000}{\includegraphics{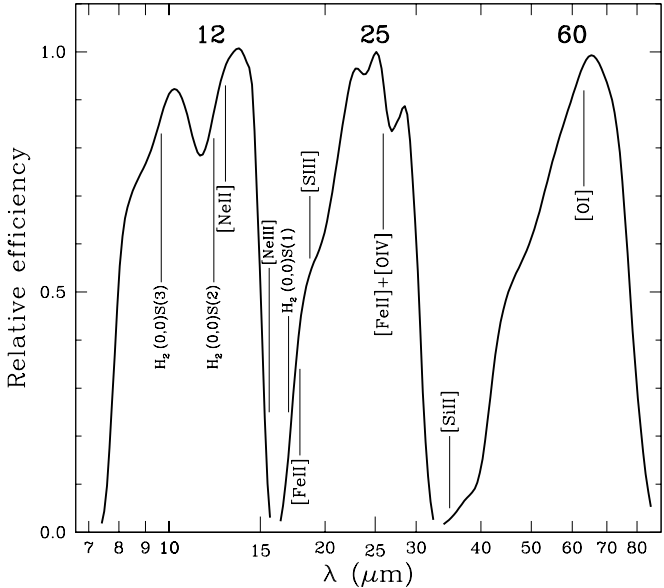}}}}
%\vskip-5pt
\caption{
Instrumental spectral response of IRAS (adapted from Fig. 2 of Neugebauer
et al.  \cite{neugebauer84}) with overlaid the positions of the brightest
ionic, atomic and molecular lines.
}
\label{iras_filt}
\end{figure}

The positions of the most important lines are marked in Fig.~\ref{iras_filt}
where one can see that [SiII]\L34.8, although very prominent in SNRs, 
does not contribute significantly because it falls at a wavelength where
IRAS was virtually blind. 
The same applies to [NeIII]\L15.6 which has an intensity comparable to
[NeII] but falls in the hole between the 12 and 25 \mic\   
filters.\\

\begin{table}
\caption{Observed ISO line fluxes }
\label{tab_flux}
\def\SKIP#1{\noalign{\vskip#1pt}}
\def\P{$\,$(}
\def\UNC{\rlap{:}}
\def\UNO{\rlap{$^{(1)}$}}
\def\DUE{\rlap{$^{(2)}$}}
\def\TRE{\rlap{$^{(3)}$}}
\def\QUA{\rlap{$^{(4)}$}}
\def\NB{\rlap{$^a$}}
\def\NC{\rlap{$^b$}}
\def\ND{\rlap{$^c$}}
\def\NE{\rlap{$^d$}}
\def\X{ $\times$ }
\def\SB{$\rm [$}
\begin{flushleft}
\begin{tabular}{lccccc}
\hline\hline
\SKIP{1}
 Line & \hglue00pt Flux\UNO & 
            \hglue10pt   Slit\DUE\hglue10pt\  &  
\multicolumn{3}{c}{$\Sigma$--IRAS\TRE}\\
   &  &  & 12 \mic & 25 \mic & 60 \mic \\
\SKIP{2}
\hline
\SKIP{1}
\SB NeII]\L12.81    & 18\P3)  & 14\X27 & 20 & -- & --  \\
\SB NeIII]\L15.56   & 10\P2)  & 14\X27 & -- & -- & --  \\
\SB FeII]\L17.93    &  4\P1) & 14\X27 & -- & 2 & --  \\ 
\SB SIII]\L18.71    &  4\P2)  & 14\X27 & -- & 3 &  -- \\ 
\SB OIV]\L25.88     &  3\P1)   & 14\X27 & -- & 3 &  -- \\
\SB FeII]\L25.98    &   8\P1)  & 14\X27 & -- & 10 &  -- \\ 
%\SB SIII]\L33.47    &   9\P3)   & 20\X33 & -- & -- & -- \\  
\SB SiII]\L34.8     & 52\P7) & 20\X33 & -- & -- & 2 \\
\SKIP{10}
\multicolumn{3}{c}{Observed IRAS fluxes$^{a}$}
                                     & 10 & 10 & 60 \\
%
% Equivalent values in Jy/sr          7.4e5 1.9e6 3.9e6
% Eq. cont-flux in SWS apert. (Jy)    0.007 0.02  0.03
\SKIP{2}
\hline
\SKIP{2}
\end{tabular}
\def\NOTA#1#2{
\hbox{\vbox{\hbox{\hsize=0.030\hsize\vtop{\centerline{#1}}}}
      \vbox{\hbox{\hsize=0.97\hsize\vtop{\baselineskip2pt #2}}}}\vskip2pt}
\NOTA{ $^{(1)}$ }{ Observed line flux, units of $10^{-16}$ W m$^{-2}$,
errors are given in parenthesis }
\NOTA{ $^{(2)}$ }{ Size (in arcsec) of the SWS apertures }
\NOTA{ $^{(3)}$ }{ Equivalent surface brightness in the IRAS bands,
units of $10^{-8}$ W m$^{-2}$ sr$^{-1}$
}
\NOTA{$^a$}{Observed IRAS surface brightness at the position
of the SWS aperture, estimated from Fig.~30a of \cite{A89},
see also Fig.~\ref{iras_map}. } 

\end{flushleft}
\end{table}
%
% Modified
%
The possible contribution of [OI]\L63.0 to the IRAS 60 \mic\  band
was already pointed out
by Burton et al. (\cite{burton90}) who observed this line in the
southern (``molecular'') rim of IC443
using a 33" aperture spectrometer on the KAO and
found line surface brightnesses a factor of $\sim$2 larger than those 
measured by IRAS.
%
% Added
%
%It should be noted, however, that the region mapped by Burton et al.
%has an emission spectrum dominated by neutral and molecular species
%and very different than the optical filaments in the NE rim where
%[OI] observations are not available.
On the other hand, however,
ISO--LWS measurements of [OI]\L63.0 through a $\oslash$ 80" aperture
centered on the optical filaments of W44
 (Reach \& Rho \cite{reach}) 
yield a line surface
brightness a factor of $\simeq$2.5 {\it lower} than the IRAS 60 \mic\  
flux\footnote{Note that the ``continuum'' in the
LWS spectra of W44 by Reach \& Rho (\cite{reach}) is most 
probably an instrumental artifact, its level being 10 times brighter
than the IRAS 60 \mic\  background+source flux}
indicating, therefore, that the contamination by [OI] is relatively 
unimportant, at least in this remnant.\\

The H$_2$ rotational lines could strongly affect the 12 \mic\  
band and may explain why the S(12)/S(25) ratio is
a factor of $\simeq$2.5 larger 
on the southern rim of IC443, i.e. at $\simeq$ 06$^h$14$^m$40$^s$ +22$^o$23' 
(cf. Fig.~\ref{iras_map}).
This region is a well known powerful source of H$_2$ (1,0)S(1)\L2.12
line emission (Burton et al. \cite{burton88}) whose spatial
distribution closely resembles the IRAS 12 \mic\  contours.
Ground--based observations
of the (0,0)S(2)\L12.3 rotational transition were obtained by
Richter et al. (\cite{richter}) who measured a line flux
$\simeq$10 times brighter than (1,0)S(1)\L2.12.
Assuming a constant I(\L12.3)/I(\L2.12) ratio over the large area mapped in
the latter transition by
Burton et al. (\cite{burton88}) yields an average line surface brightness
of about $1.5\,10^{-7}$ W m$^{-2}$ sr$^{-1}$ or 1/3 of the IRAS 12 \mic\
peak surface brightness. 
Considering that the H$_2$ spectrum of IC443 is remarkably similar to
that of Orion peak1 (e.g. Richter et al. \cite{richter}) one then expects
similar fluxes in the (0,0)S(3)\L9.66 and (0,0)S(2)\L12.3 lines
(Parmar et al. \cite{parmar}) or, equivalently, that the two lines 
should account for about 2/3 of the IRAS 12 \mic\  flux.

Finally, it should be noted that IRAS was virtually blind to the
H$_2$(0,0)S(1)\L17.0 line (cf. Fig.~\ref{iras_filt}) while the highly
forbidden ground state transition (0,0)S(0)\L28.2 is most probably too
weak to significantly contaminate the 25 \mic\  IRAS band (cf. e.g.
Fig. 2 of Oliva et al. \cite{rcw103_iso}).

%
% Main contribution are 0-0S(3) at 9.66 mic and 0-0S(2) at 12.3 mic
%
% Ricordati che F[0-0S(3)]/F[1-0S(1)] = 6.2e-3 exp(3091/T)
% Ricordati che F[0-0S(2)]/F[1-0S(1)] = 4.6e-4 exp(5270/T)
%

\section{ Conclusions }
\label{conclusions}

An ISO--SWS spectrum of IC443 has revealed prominent [NeII] and [FeII] 
line emission which, together with [SIII] and [OIV], account for most
of the observed IRAS flux in the 12, 25 \mic\   bands. 
Simple arguments indicate that this is probably the case in other
radiative SNRs and this result
suggests that the unusually blue IRAS colours of radiative SNRs
simply reflect line contamination, rather than a large population
of small grains which are otherwise required to explain the warmer
``continuum'' emission.
Available ground based data also indicate that S(2), S(3) rotational lines
of H$_2$ contribute to a large fraction of the IRAS 12 \mic\    
emission from the southern rim of IC443.\\

The relative fluxes of the ionic lines detected by ISO yield
a $\simeq$0.6 $\times$ solar Fe gas--phase relative abundance which is
significantly lower than that found in the much more powerful RCW103 
supernova remnant.
%
% Modified
%
This may imply
that the shock in IC443 is slower and thus less effective in destroying the
Fe--bearing grains.
This scenario is also supported by the lower line
surface brightness and [NeIII]/[NeII] ratio
 which both indicate a lower shock speed in IC443.

\begin{acknowledgements}
We thanks the referee, R. Arendt, for useful comments and criticisms.\\
E. Oliva acknowledges the partial support of the Italian Space Agency (ASI)
through the grant ARS--98--116/22.\\
SWS and the ISO
Spectrometer Data Center at MPE are supported by DLR (formerly DARA) under
grants 50--QI--8610--8 and 50--QI--9402--3.
\end{acknowledgements}

\end{document}